\documentclass[10pt]{iopart}
\usepackage{graphicx}
\usepackage{amssymb}
\usepackage{amstext}
\usepackage{bm}
\usepackage{setstack}
\usepackage{cite}
\usepackage{url}

\graphicspath{{Figures/}}

\begin{document}
\title[Weak knotting in confined, open random walks]{Knotting and weak knotting in confined, open random walks using virtual knots}

\author{Keith Alexander\textsuperscript{1}, Alexander J Taylor\textsuperscript{1},
Mark R Dennis\textsuperscript{1,2}}

\address{\textsuperscript{1}H H Wills Physics Laboratory, University of Bristol, Bristol BS8 1TL, UK}
\address{\textsuperscript{2}School of Physics and Astronomy, University of Birmingham, Birmingham, B15 2TT, UK}
\ead{mark.dennis@physics.org}
\begin{abstract}
We probe the character of knotting in open, confined polymers, assigning knot types to open curves by identifying their projections as virtual knots.
In this sense, virtual knots are transitional, lying in between classical knot types, which are useful to classify the ambiguous nature of knotting in open curves.
Modelling confined polymers using both lattice walks and ideal chains, we find an ensemble of random, tangled open curves whose knotting is not dominated by any single knot type, a behaviour we call weakly knotted.
We compare cubically confined lattice walks and spherically confined ideal chains, finding the weak knotting probability in both families is quite similar and growing with length, despite the overall knotting probability being quite different.
In contrast, the probability of weak knotting in unconfined walks is small at all lengths investigated.
For spherically confined ideal chains, weak knotting is strongly correlated with the degree of confinement but is almost entirely independent of length.
For ideal chains confined to tubes and slits, weak knotting is correlated with an adjusted degree of confinement, again with length having negligible effect.
\end{abstract}
\noindent{\it Keywords\/ Confined random walks, knot recognition, knot statistics, open knots}

\submitto{\jpa}

\section{Introduction}
\label{sec:int}

Random walks are natural models of flexible polymers, and statistical quantities can be probed by their computer simulation where experiments would be challenging~\cite{flory53,rubinstein03}.
An intriguing feature of polymers, and the random walks which model them, is the possibility they might be knotted~\cite{tezuka02, orlandini07, orlandini17a, orlandini17b}.
It has long been established that as the chain length of polymers increases, chains which do not contain a knot become exponentially rare, and the knots which do occur become increasingly complex~\cite{sumners88, pippenger89}.
However, length is not the sole property which affects the probability of knotting.
Unsurprisingly, polymers which are compact (i.e.~enclosed within small volumes), whether through solvent conditions or external confinement, are also more likely to be knotted, with  knotting more likely to be complex~\cite{van90, tesi94, arsuaga02, micheletti06, micheletti08, tubiana11b, poier14}.
The presence of knots in these systems has physical consequences, for example in the ejection of viral DNA from a bacteriophage capsid~\cite{marenduzzo13}, or in merely restricting the available space of conformations which can be explored in a given window of time.
This has potential ramifications for microfluidic systems which use a variety of confined environments~\cite{whitesides06, theberge10, sackmann14}.

Many previous investigations of this knot proliferation have dealt with ring polymers~\cite{tesi94, grosberg96, marcone07, orlandini07, mansfield10, tubiana11b, dai12, micheletti12a, micheletti12b}.
This has the advantage that any knots in the polymer are invariant (except through cutting and glueing processes such as the action of DNA topoisomerases and recombinases).
However, most physical polymers are linear and open-ended, which presents a problem, as knots are only defined mathematically in closed curves, up to ambient isotopy.
While an open curve is always topologically trivial, its spatial conformation can possess geometric resemblance to a knotted closed curve: the difference between an untied and tied shoelace is evident from inspection.
If the endpoints of the polymer are on strands extended away from its bulk, it is relatively easy to close the ends unambiguously, trapping the knotted area for identification.
However, when the endpoints lie within the polymer tangle (which is especially common in compact curves), this operation is much more difficult to define without ambiguity; there might be several possible knot types captured depending on the chosen closure path through the tangle~\cite{tubiana11a}.
Therefore there is an inherent ambiguity in determining whether an open curve may be described as knotted, and if so, identifying the type of `open knot'.

In such situations, a standard method is to take many closures according to a statistical rule, and look for the most common knot type; this might involve  extending the endpoints on straight lines to a point on the surface of a large sphere enclosing and centred on the curve, thus giving a closed curve with a definite knot type for each point on the sphere~\cite{millett05a, millett05b}.
We call this method \emph{sphere closure}.
Upon averaging over area on the sphere closures, usually the most common knot---i.e.~the modal average---is taken to represent the knot type of the open curve.

Complementing this approach, we presented an alternative closure method based on identifying projections of the open curve as \emph{virtual knots}~\cite{alexander17}, a generalization of the familiar, classical knots which can be interpreted as lying `in-between' the standard knot types~\cite{kauffman99}.
Our method, which we call \emph{virtual closure}, involves projecting the open 3D space curve to an open knot diagram, and we consider projections in all directions over the sphere.
The open knot diagrams are interpreted either as non-classical virtual knots, or classical knots; in the following, to avoid confusion, we refer to a knot only as virtual if it is not of a classical type, and these non-classical virtual knots types have been partially classified~\cite{virtualknottable}.
By taking many projections from uniform directions, we again build up a picture of the knotting of the open curve.
Including the possibility that an open curve projection is identified as a virtual knot increases the sensitivity to identifying how an open space curve may be knotted---quantifying the inherent ambiguity---with a subtlety missed by sphere closure.
In Ref.~\cite{alexander17} we mainly applied this method to knotted proteins; here we extend it to more general open random walks.

Recently, consideration of open curves has led to the study of \emph{knotoids}~\cite{goundaroulis17a}.
These are essentially open knot diagrams of the sort that result from projections of open curves, and they behave very similarly to the virtual knots which occur as open knot diagrams.
Indeed, while knotoids in principle can distinguish open curve projections better than virtual knots, in practice suitably powerful knotoid invariants have not been used to our knowledge.
By considering knotoid diagrams as drawn in the plane (rather than the surface of a sphere, like virtual knot diagrams), open curve projections can be further distinguished with a simple extension of the Jones polynomial~\cite{goundaroulis17b}.
The important common feature of virtual closure and knotoid methods is that the problem of open curve knot recognition is mapped to distinguishing ensembles of open knot diagrams, instead of ensembles of closed curves as in sphere closure.

In trying to understand how an open curve might be identified as (virtually) knotted, we consider the statistics of the knot types that occur over different projections.
As mentioned above, in earlier sphere closure studies, an open curve is identified with the most common knot type occurring over different closures, even if this does not cover over 50\% of closures.
For example, the trivial unknot may be the single most common knot type, occurring in 40\% of closures.
Should we say this curve is unknotted, despite 60\% of closures being knotted (albeit in different ways)?

We attempted to resolve this in~\cite{alexander17} by \emph{defining} an open space curve to be knotted if at least 50\% of closures result in a non-trivial knot (classical or virtual).
Situations where the single most common knot type is non-trivial and covers at least 50\% of closures we call \emph{strongly knotted}.
This leaves situations where the most common knot type occurs in less than 50\% of closure directions, interpreted as significant ambiguity in the knot type of the open curve.
Such space curves are called \emph{weakly knotted}: they are definitely tangled, but not in a way that can be recognised as a single knot type, classical or virtual.
It is weak knotting that confounds the classification of knotting in open curves.

Unconfined, open polymers typically appear as space curves whose endpoints are extended away from the bulk~\cite{grosberg96, bao03}, and are therefore likely to be either unknotted or strongly knotted.
Knotting ambiguity is not of much concern in such systems, as observed in early studies of sphere closure, although not quite in these terms~\cite{millett05a}.
On the other hand, \emph{confined, open polymers} tend to have more compact and complex conformations, and we expect each curve to project to different knot types in different directions, making these good candidates for weak knotting.
There have already been indications of this from a study comparing knot recognition methods~\cite{tubiana11a}, in which the different methods tended to disagree on knot type increasingly often with tighter confinement.

While we do not investigate it directly here, there is an important link between weak knotting and shallow knotting, in which the knot type of an open curve can be changed (often to the unknot), by removing a short length from one of the endpoints.
Such shallow knotted configurations are common in protein knots and are also present as transition states in polymer dynamics, for example during an untying process, making weak knotting an important indicator of this.
We expect virtual closure to be a better detector of weak knotting than sphere closure, as knot projections might be identified as virtual as well as classical knot types.

\begin{figure}
  \centering
  \includegraphics[width=\textwidth]{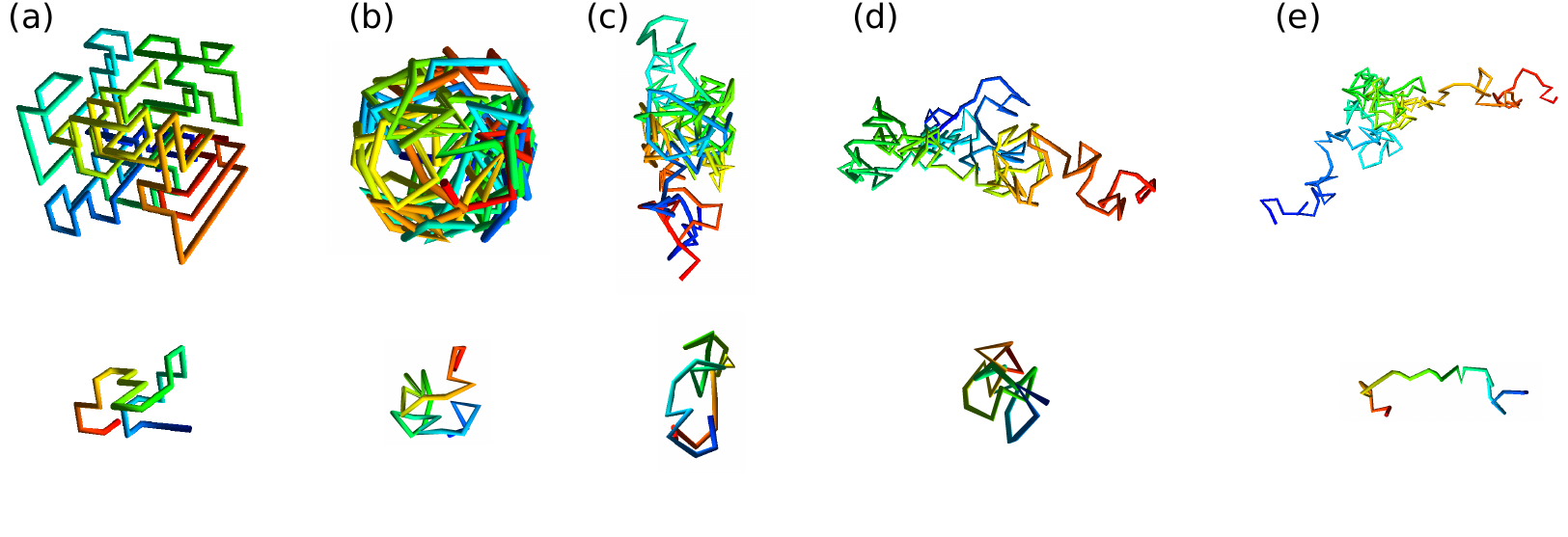}
  \caption{Examples of the classes of open random walks investigated.
  (a) Cubically confined lattice walks.
  Off-lattice walks (b) confined to spheres; (c) confined to tubes; (d) confined to slits; (e) unconfined.}
  \label{rw_examples}
\end{figure}

Here we quantitatively investigate how confinement affects the weak knotting of open random walks modelling confined open polymers.
We generate random walks of three different types: cubically confined lattice walks, spherically-confined off-lattice walks and unconfined off-lattice walks, and compare the results for each.
We focus our attention on spherically-confined off-lattice walks, and then generalise to different confining geometries, namely tubes and slits.
We will see that while overall knotting probability is highly dependent on walk length---as is well known for ring polymers and closed random walks---this is not the case for weak knotting.
Instead, the walk's degree of confinement, determined by the relative radius of gyration of unconfined and confined walks of the same length, determines the weak knotting probability with no further contribution from length.
Some examples of random walks in these different classes are shown in Fig.~\ref{rw_examples}.

\section{Classical and virtual knotting of open walk closures}
\label{sec:cv}

As in our survey of virtually knotted proteins~\cite{alexander17}, we employ both sphere closure and virtual closure to analyse the knotting of open random walks.
For each open chain, we consider 100 choices of closure direction, uniformly distributed over the sphere of directions according to the generalised spiral points algorithm of~\cite{rakhmanov94}.
In sphere closure, we project the space curve into the plane perpendicular to the projection direction, keeping crossing information.
We then join the ends of the projections, adding over-crossings where the closure arc crosses existing strands of the diagram, equivalent to joining the ends to a point on a sphere of arbitrarily large radius.
Any projection of this closed curve, with crossings signed appropriately, gives a closed knot diagram of the same classical knot type.
We find this type by calculating the Alexander polynomial of the knot diagram~\cite{alexander28,orlandini07}.
While this may not be the strongest knot invariant, it distinguishes enough of the different knot types that occur in the analysis of a single curve that we can reliably judge the weakness of knotting.
With a perfect knot invariant, we would expect to see a very marginal increase in weak knotting detections.

Under virtual closure, the endpoints of the projected diagram are joined making \emph{virtual crossings} with existing strands, which are not `classical' over- or under-crossings, and indeed should not be interpreted as physical crossings.
This procedure gives a closed (virtual) knot diagram whose topological information is only that of the classical crossings in the original projection.
We use the generalised Alexander polynomial to distinguish virtual knot types~\cite{kauffman03, sawollek99}.
As the generalised Alexander polynomial cannot distinguish certain simple virtual knots, we also employ the Jones polynomial when necessary~\cite{jones85}.
All classical knot types (including those with virtual crossings removable by virtual Reidemeister moves) have a generalised Alexander polynomial of zero; the classical Alexander polynomial then distinguishes these.

\begin{figure}
  \centering
  \includegraphics[width=\textwidth]{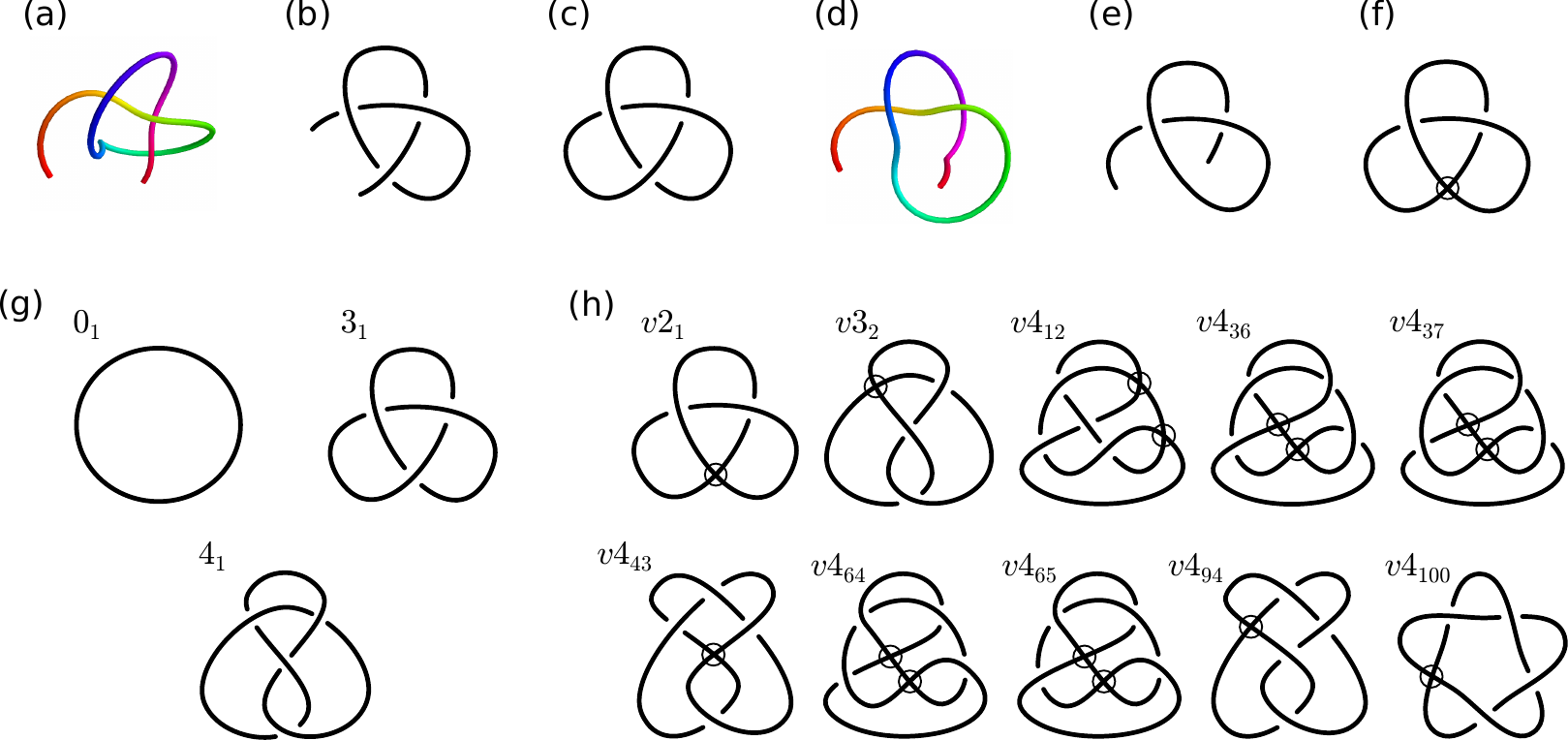}
  \caption{Examples of classical and virtual knot projections.
  (a) represents an open space curve, (b) its projected diagram and (c) the resulting closed knot diagram upon closing the endpoints unambiguously.
  (d) represents the same curve viewed from a different direction, (e) its projected diagram and (f) the resulting closed knot diagram on virtual closure.
  Sphere closure (including an over-crossing) results in the same knot as (c).
  (g) gives standard projections of the three classical knots of up to four crossings: the topologically trivial unknot, $0_1$, the trefoil knot $3_1$ and the figure-8 knot $4_1$.
  (h) shows virtual knots possible from projections of open curves of up to four crossings, using the notation of \cite{virtualknottable}.  
  }
  \label{virtual_knots}
\end{figure}

Examples are useful to illustrate the different closure mechanisms.
Fig.~\ref{virtual_knots}~(a) shows an open curve embedded in 3-space, with (b) showing the corresponding projected diagram (the diagram has been smoothed, but no crossings have changed).
Fig.~\ref{virtual_knots}~(c) shows the closure of the projected diagram.
Since no strands separate the endpoints, the closure introduces no additional crossings: both sphere and virtual closure identify this projection as trefoil knotted (i.e.~classical knot type $3_1$). 
Fig.~\ref{virtual_knots}~(d) shows the same open curve viewed from a different direction, with (e) the corresponding projected diagram.
Here, joining the strands requires an extra crossing.
Under sphere closure, this is an over-crossing, leading to a final diagram identified as a trefoil knot, as in (c).
On the other hand, making this a virtual crossing instead gives the virtual knot shown in (f), the \emph{virtual trefoil} $v2_1$. 
Identifying the projection as a virtual knot captures the fact that this diagram is intermediate between the classical unknot and the trefoil knot.

The possible classical knot types up to four crossings are shown in Fig.~\ref{virtual_knots}~(g) with their usual labels~\cite{rolfsen76}.
The full set of virtual knots includes many more members than the virtual closures of projected knot diagrams which are our interest; those virtual knots which occur under virtual closure of an open space curve, of up to four classical crossings, are shown in Fig.~\ref{virtual_knots}~(h).
The labels here are from the virtual knot table~\cite{virtualknottable}, where we add the prefix `$v$' to distinguish from the classical knots.
All of these virtual knots except $v4_{12}$ are prime virtual knots, and appeared in the classification of more general prime virtual knots of genus 1~\cite{andreevna14}.
$v4_{12}$ in fact is a composite of two virtual trefoils, $v2_1 \# v2_1$, and is not prime as stated in~\cite{alexander17}.
Notably, that composite virtual knots appear with only four crossings, unlike the six crossing minimum seen in classical knots.

Evidently, virtual closure allows for many more possible knot types than sphere closure, for a curve or its projection of the same minimum classical crossing number (which we take as a measure of complexity). 
As in the example considered above, the virtual types occur in between classical knot types, and their identification sharpens the sense of ambiguity an open-chain projection can have between classical knot types.
In particular, $v2_1$ is simpler---has fewer classical crossings---than the simplest non-trivial classical knot $3_1$, yet is still (virtually) knotted.

\section{Generating random walks}
\label{sec:generating}

We investigate both lattice and off-lattice walks.
All walks are equilateral and have a step length of 1.
In each model, for a given chain length and degree of confinement, we perform statistics based on 10,000 walks.
Our lattice walks, following \cite{lua04}, are segments of Hamiltonian walks on a cubic lattice of $L \times L \times L$ nodes, which are automatically self-avoiding.
The off-lattice walks we investigate are ideal chains, often used to model polymers in theta conditions~\cite{flory53}.
For unconfined walks, begin at the spatial origin without loss of generality and step one unit in a uniformly randomly chosen direction.
This process is repeated until the walk reaches the desired length.
Each step is independent of the walk's history, and is not self-avoiding though generically does not cross itself.

It is more complicated to generate confined ideal chains.
A simple algorithm is an accept-reject method, where the walk is generated like an unconfined ideal chain, starting at a random point within the confining volume, but whenever a step would take the walk outside the confined volume, it is rejected, and a new step sampled until a point inside the volume is returned. 
Boundaries implemented in this way are called absorbing boundaries -- vertices of walks generating this way tend to avoid the region within a step length of the boundary.

\begin{figure}
  \centering
  \includegraphics[width=\textwidth]{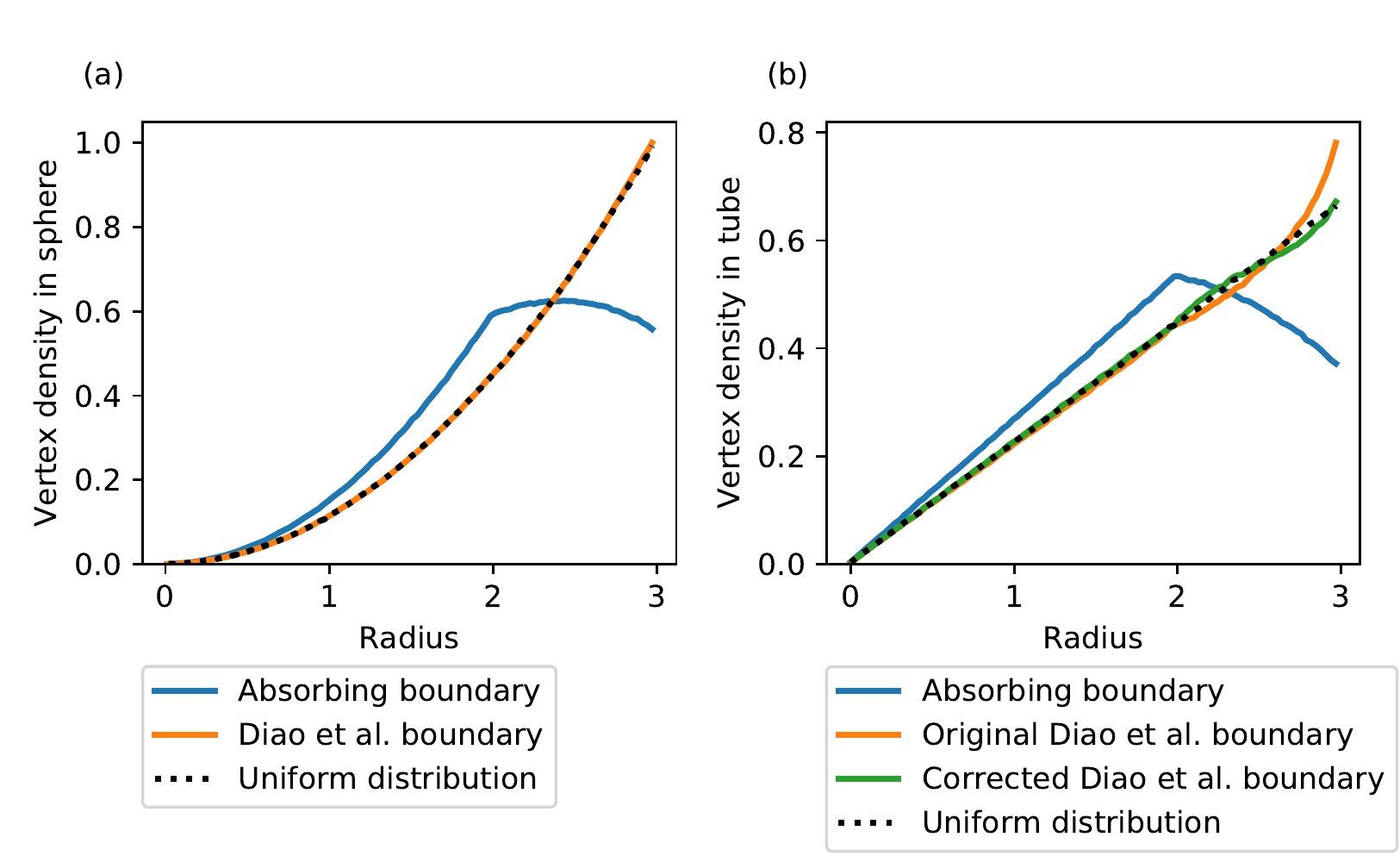}
  \caption{Distributions of vertices for spherically-confined random walks.
  (a) Radial distribution of vertices in a sphere with absorbing boundaries and boundaries using the model of \cite{diao12} compared to a uniform distribution.
  (b) Radial distribution of vertices in a tube with absorbing boundaries and boundaries using the model of \cite{diao12}, and our modification of it, compared to a uniform distribution.}
  \label{radial_distribution}
 \end{figure}

A better sampling of walks confined within a sphere, very close to a uniform distribution, was proposed in~\cite{diao12}.
Here, walks are generated like ideal chains, unless they are under a step length from the boundary, where each step is sampled from a distribution which increases towards the boundary.
The radial density of walk vertices from each method is given in Fig.~\ref{radial_distribution}~(a), compared to a uniform distribution.

It is easy to adapt this method for walks in slits by treating each parallel wall as an arbitrarily large sphere.
More difficult is adapting to tubes, which requires steps from within a step length of the boundary to be sampled from a differently shaped distribution than that of the sphere.
We outline how we do this in the Appendix.
While the results are not as close to uniform as the spherical case, they are close enough that any effect on knotting statistics is likely to be small.
The radial vertex densities of walks in tubes using absorbing boundary conditions, the original distribution shape and our corrected distribution compared to uniform are given in  Fig.~\ref{radial_distribution}~(b).

\section{Results}

 \subsection{Knotting in three-dimensionally confined random walks}

We begin by investigating the dependence of knotting probability on chain length, comparing unconfined ideal chains, spherically confined ideal chains and walks on the finite cubic lattices of $L \times L \times L$ nodes where $L = 6$ or $7$.
For the best comparison between ideal chains and lattice walks, the ideal chains are confined to spheres of equivalent volume as the lattices, with radius $R = \left(\frac{3}{4\pi}\right)^{1/3}(L-1) \approx 3.1$ and $3.72$ respectively.

\begin{figure}
  \centering
  \includegraphics[width=\textwidth]{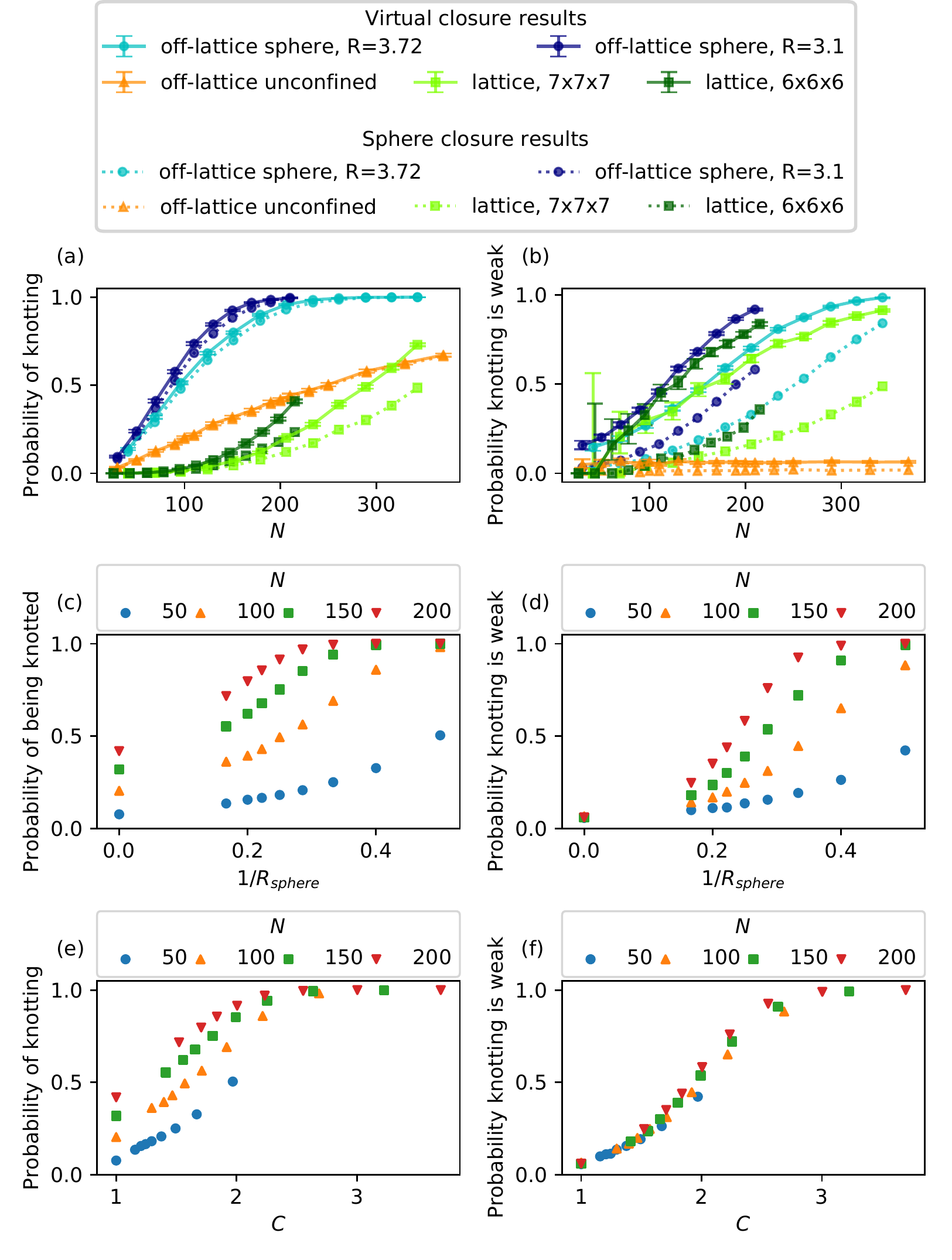}
  \caption{Knotting and weak knotting probabilities for the different random chain models. 
  Plots on the left are for knotting probability, on the right conditional probability that knotting is weak.
  (a) Knotting and (b) probability knotting is weak, with different colours representing different chains as in the key. 
  (c) and (d) show (weak) knotting in spherically confined ideal chains as the confining radius is varied.
  (e) and (f) show the same data as (c) and (d), plotted against the degree of confinement $C$ as defined in (\ref{Cdef}).}
  \label{rw_knotting_probability}
 \end{figure}

Fig.~\ref{rw_knotting_probability}~(a) shows the probability that walks of $N$ steps are knotted, i.e.~if 50\% or fewer closures are unknotted.
The solid line gives the knotting probability under virtual closure and the dashed line knotting under sphere closure.
For each walk model, knotting probability increases with length, and is strongly encouraged by confinement.
The reduced flexibility of the lattice walks lowers the knotting probability compared to the ideal chains at shorter lengths, beginning to overtake unconfined ideal chains at longer lengths.
For the ideal chains, the knotting probability is very similar under sphere closure and virtual closure.
There is a greater discrepancy in the lattice walk results, as the knots found are likely to be simpler (and hence more cases of $v2_1$ versus the unknot). 
For complex ideal chains, the two methods might disagree in the knot which occurs, but not whether it is knotted; on a lattice there are more conformations which are ``only just'' knotted.

We now examine the conditional probability that , given a knotted walk, the knotting is weak, i.e.~no single knot type occurs in 50\% or more closures.
Note that this is distinct from the probability of weak knotting; we want to capture the character of the knotted walks.
The results for the same chains are plotted in Fig.~\ref{rw_knotting_probability}~(b); unlike the knotting considered above, unconfined walks display almost no weak knotting at any length. 
On the other hand, all confined walks show an increase in weak knotting with length.
Strikingly, under virtual closure the trends for ideal chains and lattice walks, for the same volume size, are very close.
Sphere closure gives significantly less weak knotting, primarily due to the reduced number of knot types possible under sphere closure, which reduces the competition for the most common knot.
Lattice walks show the greatest reduction, where the simplicity of the curves means the number of knot types accessible is less than in the geometrically complex ideal chains.

From these results we see that confinement drives knotting from almost entirely strong knotting to almost entirely weak knotting (as a measure of knot type ambiguity).
To probe this effect directly, we consider ideal chains of fixed length confined in spheres of different radius, under virtual closure (which is more sensitive to weak knotting). The dependence of (weak) knotting of fixed-length chains with respect to reciprocal confining radius $1/R$ is shown in Fig.~\ref{rw_knotting_probability}~(c) and (d), where the unconfined walks are at zero. 
For all lengths investigated the knotting probabilities increase as the confining volume decreases until the probability saturates, with longer chains more likely to be knotted.
As expected, with tighter confinement the knotting is more likely to be weak, saturating with all knotted walks being weakly knotted.

When walks are unconfined ($1/R \to 0$), the probability is small for knotting to be weak, for all walk lengths.
Now, tightening the confinement means the probability of weak knotting rises faster for longer walks, as these are more constrained by a given confinement volume.
We quantify this degree of confinement $C$ as 
\begin{equation}
   C \equiv \frac{\langle R_g(N, R=\infty)\rangle }{\langle R_g(N, R)\rangle},
   \label{Cdef}
\end{equation}
where $\langle R_g(N, R)\rangle$ is the average radius of gyration of the walk of length $N$ confined to a sphere of radius $R$ ($\langle R_g(N, R=\infty)\rangle$ is the average radius of gyration of unconfined walks of length $N$). 
We expect $C \ge 1$, with $C = 1$ suggesting no confinement.

The probability of knotting with respect to $C$ is shown in Fig.~\ref{rw_knotting_probability}~(e).
Clearly knotting probability increases with $C$, and also with chain length.
However, this length dependence does not seem to be present for the probability knotting is weak against $C$, plotted in Fig.~\ref{rw_knotting_probability}~(f) -- the points all lie close to the same curve, over the whole range from no weak knotting at $C = 1$ to a probability of unity at $C \approx 3$.
Thus the degree of confinement as parametrised by $C$ has a crucial correlation with weak knotting, with unconfined walks not exhibiting weak knotting, and strongly confined knots almost surely weakly knotted, with the most common (virtual) knot type not covering a majority of closures.

\begin{figure}
 \centering
 \includegraphics[width=\textwidth]{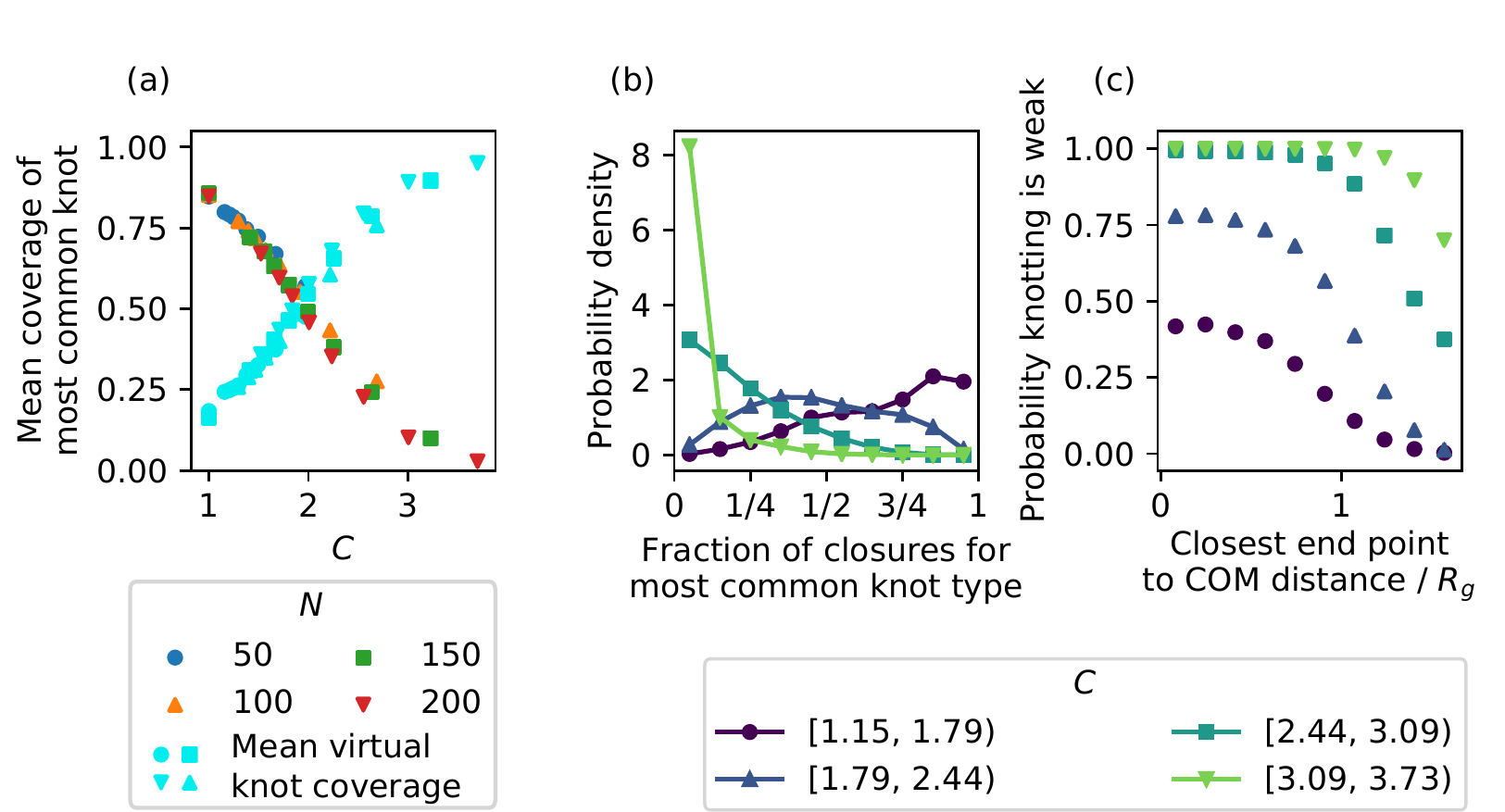}
  \caption{Further results for the virtual closure analysis of spherically confined ideal chains.
  (a) Mean coverage of the most common knot against the degree of confinement $C$ (see (\ref{Cdef})).
  In cyan is shown the mean coverage of virtual knots for the same walks.
  (b) The probability distribution of fraction of closures of the commonest knot type.  
  Results for walks have been binned according to $C$. 
  (c) Probability knotting is weak against the displacement from the centroid of the closer endpoint, divided by radius of gyration.
  Again, walks are binned by degree of confinement. }
  \label{most_common_knot_and_end_position}
\end{figure}

Our definition of (weak) knotting places great emphasis on using 50\% as a cutoff.
However, it is unlikely that curves whose most common knot type covers 51\% of closures and those with only 49\% coverage are very different conformationally.
For walks of a given length and confining radius, a more natural question is, what is the mean coverage of the most common knot?
This proportion of closure directions covered by the most common knot for knotted curves is plotted against $C$ in Fig.~\ref{most_common_knot_and_end_position}~(a).
Again, the dependence on $C$ appears to be effectively independent of chain length, with a mean coverage of almost the whole sphere for $C = 1$ (unconfined), decreasing to about 25\% at $C = 2.5$. 
The precision in the results for larger $C$ is limited by the number of closures (100 here) considered for each space curve, as in most cases there is a different knot type for each closure direction.
This supports the previous result that, as the degree of confinement of a chain increases, the knotting is more likely to be weak, as the most common knot type drops below 50\% at $C \approx 2$. 
Plotted also is the coverage of virtual knots with $C$, which shows a similar $N$ independence, rising as the mean coverage of the commonest knot falls.
Weaker knotting implies that a higher fraction of closures yields virtual knots.

More insight can be gained by considering the distribution of the coverage fraction of the most common knot, plotted in Fig.~\ref{most_common_knot_and_end_position}~(b), where the different curves represent walks with a similar value of $C$.
For walks with a low degree of confinement ($C \lesssim 1.8$)  the distribution is weighted towards a large fractional coverage, i.e.~knotting is often very strong.
As $C$ increases, the distribution shifts to lower coverage fractions, such that for most tightly confined walks ($C \gtrsim 3$) the distribution is sharply peaked at low fractional coverage.
It is easy to see how plot (a) results from the means of the distributions in (b).

In the Introduction, we stated that weak knotting mainly stems from the endpoints of the open random chains being close to or inside the tangled bulk of the curve.
The plot in Fig.~\ref{most_common_knot_and_end_position}~(c) verifies this statement, where the probability knotting is weak is plotted against the smaller displacement of the two endpoints from the walk centroid, normalised with respect to radius of gyration, once again binning according to $C$.
For each value of $C$, the probability knotting is weak is constant up to a distance of about $0.7 R_g$ (though dependent on $C$), then decreases to $0$, with the distance the weak knotting probability reaches zero being dependent on $C$.
An end need only be buried slightly within a very complex walk to be weak.
Of course, the radius of gyration of a strongly spherically confined walk is fixed at the confining radius.

\subsection{Knotting in off-lattice walks confined to tubes and slits}
 
\begin{figure}
  \centering
  \includegraphics[width=\textwidth]{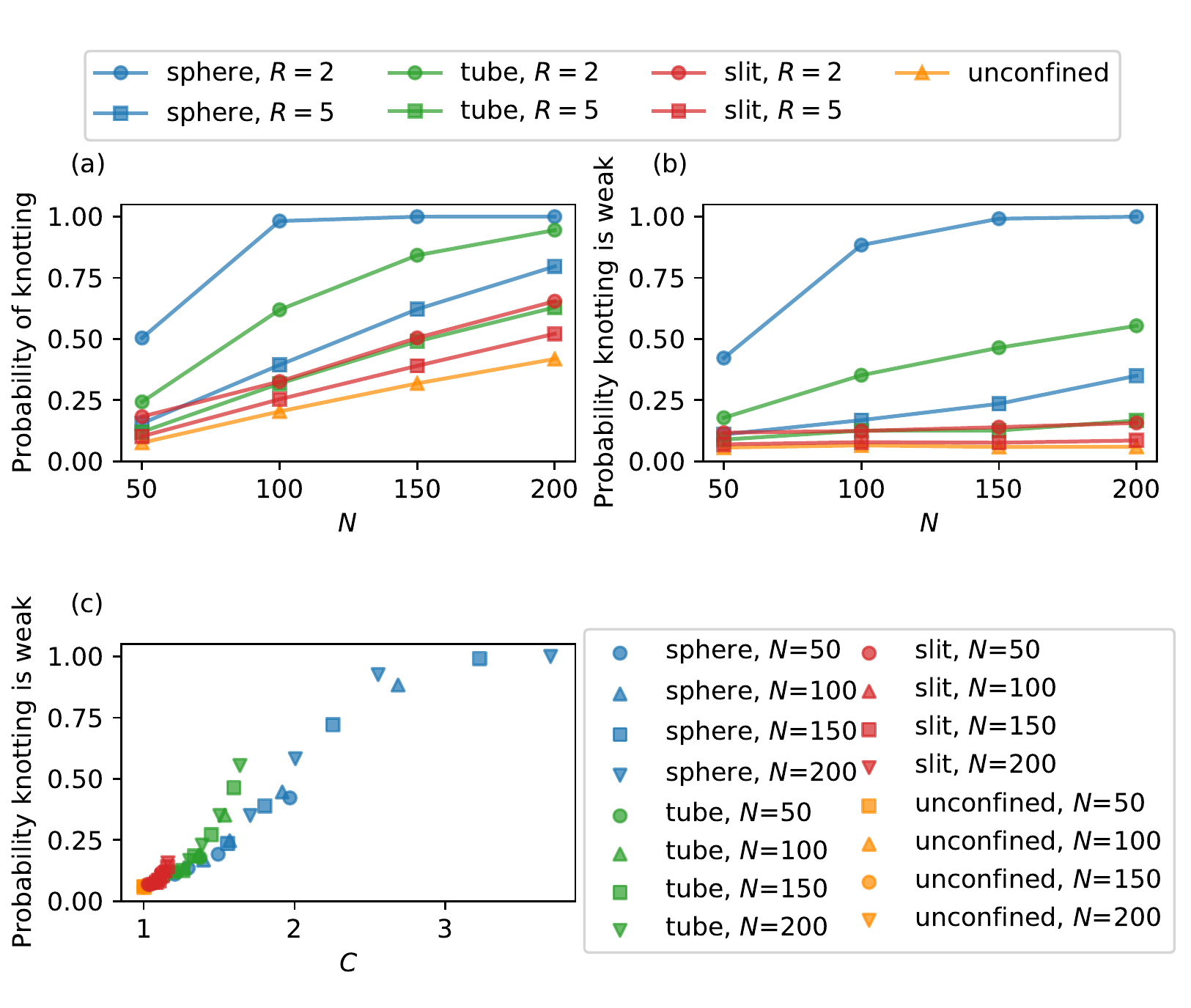}
  \caption{Knotting and weak knotting in random walks confined to spheres, tubes, slits and unconfined.
  Data for $R=2$ and $R=5$ are shown, exemplifying tight and loose confinement.
  (a) Probability of knotting against chain length.
  (b) Probability of weak knotting against chain length.
  (c) Probability of weak knotting against $C$, the degree of confinement relative to the radius of gyration.}
  \label{compare_confinements_knotting}
\end{figure}

We now investigate the effect of different forms of confinement on knotting and weak knotting, varying the number of confined dimensions from three in the sphere, to two in the tube, one in the slit and none when unconfined.
The knotting probability against chain length for these different cases is plotted in Fig.~\ref{compare_confinements_knotting}~(a), for two choices of confining radius, 2 (relatively tight) and 5 (loose). 
The radius of slits is half the distance between the walls.
As before, knotting probability increases with chain length, and as the confining radius decreases. 
There are fewer knots in tubes compared to spheres of equal radius and fewer yet in slits, as we would expect.
Changing the confining radius has less effect on knotting in lower-dimensional confinement: there is little change in slits and less effect in tubes than spheres.
We do not investigate sufficiently tight tubes and slits to see the non-monotonic change in knot probability found in \cite{orlandini13}.

Knotting is much more commonly weak in spheres than tubes and slits, shown in Fig.~\ref{compare_confinements_knotting}~(b).
The growth of weak knotting with length, compared to the previously seen effect in spheres, is modest in tubes and almost non-existent in slits, and the effect of tightening the confining radius in slits is again negligible.

These results should not be surprising since confinement in one and two dimensions, realised by slits and tubes respectively, has a smaller impact than confinement in three dimensions in spheres.
To see this directly, we plot the probability of weak knotting in these partially confined geometries against the degree of confinement $C$ (defined in (\ref{Cdef})) in Fig.~\ref{compare_confinements_knotting}~(c).
We investigate walks in tubes and slits over a much smaller range of degree of confinement than previously for spheres; nevertheless, the data indicates that probability knotting is weak against $C$ is independent of confinement radius and walk length for both tube confinement and slit confinement, with knotting more likely to be weak, for fixed $C$, for slit confinement than tube, and tube than sphere.
This is because partial confinement has a more significant effect on weak knotting than on the radius of gyration with respect to which it is normalised.

The radius of gyration fails to quantify the anisotropy of partially confined walks: compared to the prolate, cigar-like shape expected for unconfined walks, walks in slits tend to be less prolate and walks in tubes more prolate. 
We therefore consider the rank two, symmetric gyration tensor $\mathbf{S}$ with components $S_{ij}$, 
\begin{equation}
  S_{ij} = \frac{1}{N}\ \sum_{n=1}^N (r_i^{(n)} - \overline{r}_i)(r_j^{(n)} - \overline{r}_j), \qquad i,j = 1,2,3, 
  \label{eq:gyr_tensor}
\end{equation}
where $\bi{r}^{(n)}$ is the $n$th walk vertex, $n = 1, \ldots, N$, and $\overline{\bi{r}} = N^{-1} \sum_{n=1}^N \bi{r}^{(n)}$ is the walk centroid.
With this terminology, the radius of gyration $R_g = \sqrt{\tr \mathbf{S}}$.
A suitable orthogonal transformation $\mathbf{O}$ (i.e.~choice of Cartesian coordinate system) diagonalises this tensor,
 \begin{equation}
  \centering
  \mathbf{O}\mathbf{S}\mathbf{O}^T =  \rm{diag}[\lambda_1, \lambda_2, \lambda_3],
  \label{eq:gyr_tensor_eigenvalues}
\end{equation}
where the axes are ordered so $0<\lambda_1 \leq \lambda_2 \leq \lambda_3$, and clearly $R_g^2 = \lambda_1+ \lambda_2 + \lambda_3$.
Therefore $\alpha_j = \sqrt{3\lambda_j}$, $j = 1,2,3$ are the semi-axis lengths of the associated \emph{ellipsoid of inertia}, indicating the shape of the walk (with notation following \cite{rawdon08b}, where the $\alpha_i$ were used to characterise the shape of knotted ring polymers).

\begin{figure}
  \centering
  \includegraphics[width=\textwidth]{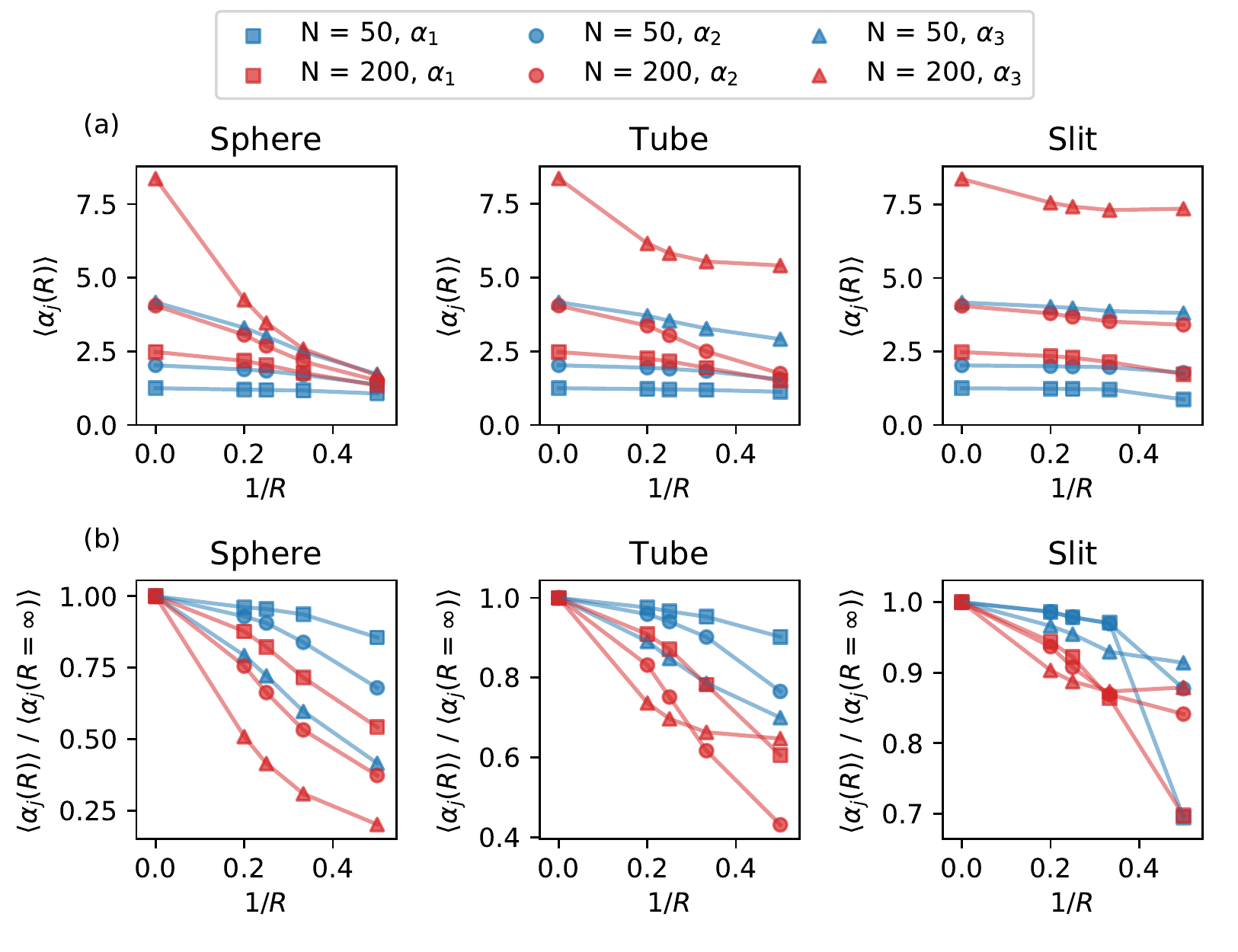}
  \caption{The variation of semi-axes of the inertia ellipsoid of open walks confined to spheres, tubes and slits.
  (a) The absolute mean size of each semi-axis in confined walks of length $N = 50$ and $200$.
  (b) The mean size of each semi-axis in confined walks, relative to unconfined walks. }
  \label{semi_axis_sizes}
\end{figure}

The mean absolute value of each $\alpha_j$ against $1/R$ for each kind of confined walk is plotted in Fig.~\ref{semi_axis_sizes}~(a).
All walks start from the prolate shape of unconfined walks at $1/R \to 0$.
The confinement is isotropic in spheres, so all $\alpha_j$ are distributed like $R_g$, which for longer walks tends to the confining radius $R$.
In tubes, $\alpha_1$ and $\alpha_2$ approach $R$, while the effect on the largest semi-axis $\alpha_3$ is more limited.
In slits, each semi-axis reduces as $1/R$ increases, but not to a great extent (unsurprising as, for the chains considered, $\alpha_1$ is often shorter than half the distance between the walls).

We also plot the relative size of the mean confined semi-axes compared to their mean unconfined sizes in Fig.~\ref{semi_axis_sizes}~(b).
In spheres, as expected, the reduction with $1/R$ is largest for the largest semi-axis $\alpha_3$.
However, in tubes for the longest walks, we find that $\alpha_3$ is the least relatively reduced, despite being the most reduced in absolute terms.
We see this inversion in slits also in both long and short walks.

This explains why the range of degree of confinement $C$ (defined in (\ref{Cdef})) is so much smaller in tubes and slits than in spheres.  
The radius of gyration is derived from the sum of squares of the semi-axis lengths, and, unconfined, we expect walks shaped such that $\alpha_3 \gg \alpha_1, \alpha_2$, therefore the radius of gyration is dominated by $\alpha_3$.
In spheres, $\alpha_3$ greatly decreases with radius, and so $C$ varies considerably.
In tubes and slits, $\alpha_3$ is less affected by the confinement than $\alpha_1$ and $\alpha_2$, despite being somewhat reduced.
This detail is lost in the radius of gyration, suggesting we may get a better predictor of weak knotting if we put each semi-axis on the same footing.
We propose the following \emph{adjusted degree of confinement}, $C^\star$:
 \begin{equation}
  C^\star = \frac{1}{3}\sum_{i=1}^3 \frac{\langle\alpha_i(N, R=\infty)\rangle}{\langle\alpha_i(N, R)\rangle},
  \label{eq:Cstardef}
 \end{equation}
summing the mean size of each semi-axis relative to those of unconfined walks, and dividing by 3 to ensure $C^\star = 1$ for unconfined walks.

\begin{figure}
 \centering
 \includegraphics[width=\textwidth]{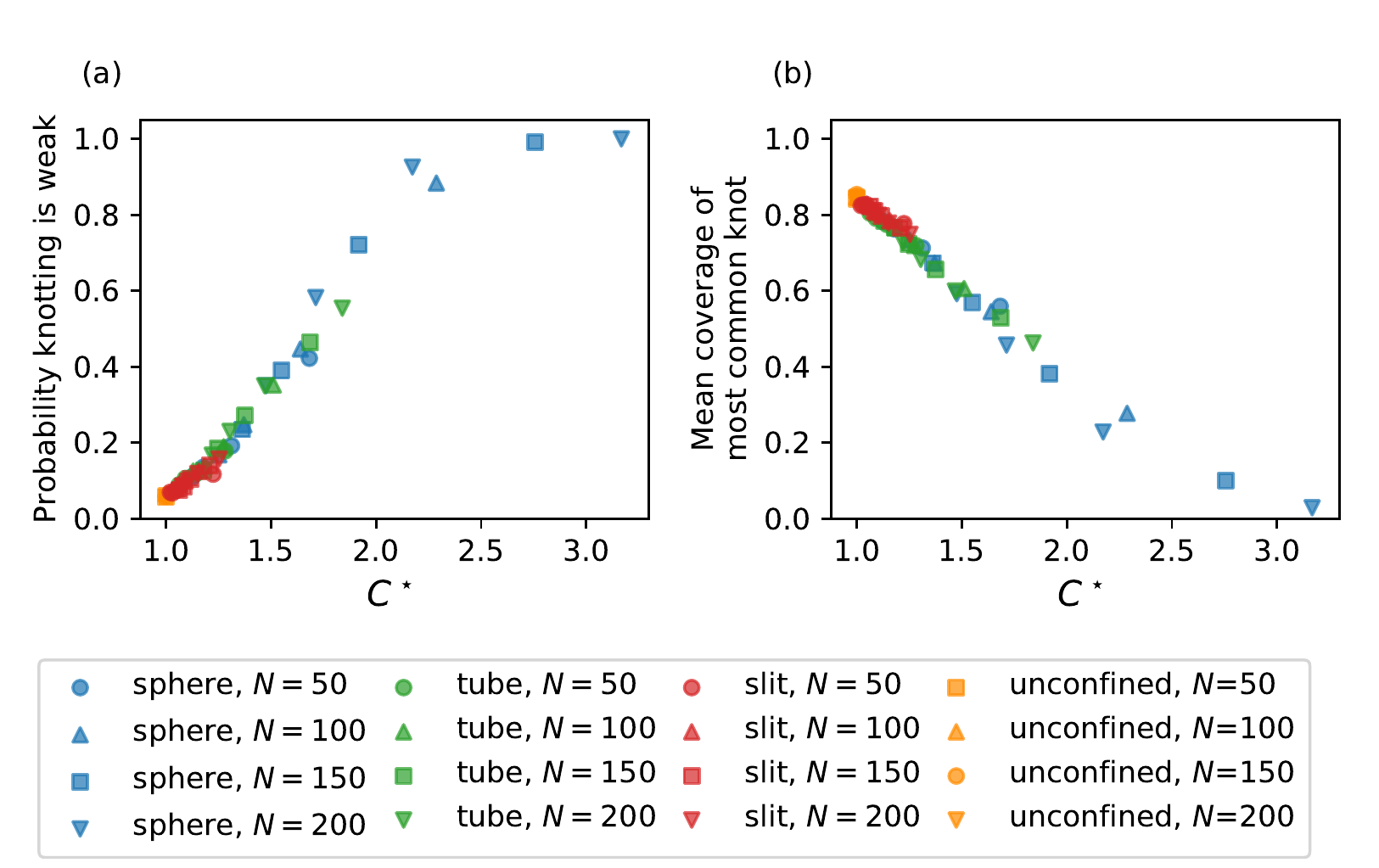}
 \caption{(a) probability that knotting is weak and (b) the mean coverage of the most common knot in ideal chains confined to different geometries, plotted against the adjusted degree of confinement, $C^\star$ (as in (\ref{eq:Cstardef})).}
 \label{adjusted_degree_of_confinement}
\end{figure}

The probability knotting is weak against this adjusted degree of confinement for walks in each confinement shape is plotted in Fig.~\ref{adjusted_degree_of_confinement}~(a).
We see that now all the values fall onto the same curve, regardless of the shape of confinement.
Spheres probe the largest range of adjusted degree of confinement, followed by tubes and lastly by slits.
Knotting is more likely to be weak for larger $C^\star$.
Furthermore, we plot the mean coverage of the most common knot against the adjusted degree of confinement in Fig.~\ref{adjusted_degree_of_confinement}~(b).
We see again that all the values for all walks lie on the same curve.
With larger $C^\star$, the mean coverage of the most common knot is lower, i.e. the weaker the knotting.
In summary, $C^\star$ is strongly correlated with the weak character of knotting in confined walks, without further dependency on length of walk, or shape or radius of confinement.

\section{Discussion}

We have performed a detailed numerical investigation into the cases when assigning a knot type to an open tangled curve is ambiguous, which we call `weak knotting'.
While unconfined open random walks can often be assigned a clear knot type, as walks are confined more and more tightly (i.e.~within a progressively smaller radius), closures in different directions become assigned to more different knot types, with no one type dominant.
We were able to quantify this ambiguity using a knot recognition method which used multiple closures to give a detailed picture of knotting.
Identifying projections using virtual knots is more sensitive to knotting, and to the ambiguity of knotting over previous sphere closure schemes.
Lattice and off-lattice walks were investigated, and while each showed different overall knotting, with lattice walks less likely to be knotted, the ambiguity in their knotting under confinement was very similar when measured using virtual closure.

In particular, we found that the relative radius of gyration of unconfined walks to confined walks is well correlated with the probability of weak knotting, independent of chain length.
We also found that weak knotting is strongly correlated to walk endpoints being close to the walk centroid, i.e.~within the bulk of the tangle.
In other words, the knotting spectrum of open random walks grows more complex with confinement, and when their endpoints are more buried in the tangle.
By varying the confining geometry, we found that a better predictor of weak knotting is using the relative reduction of each semi-axis of the characteristic ellipsoid of unconfined and confined walks.

This analysis shows that the distinction between weak and strong knotting is important when open chains are randomly tangled and confined within volumes smaller than the unconfined radius of gyration.
The methods introduced here should be useful in more extensive statistical analyses of the shapes of open macromolecule chains. 
Examples where this may prove important include systems of polymers which are compact through poor solvent conditions, or external confinement due to microfluidic environments or viral capsids.

\begin{figure}
 \centering
 \includegraphics[width=0.7\textwidth]{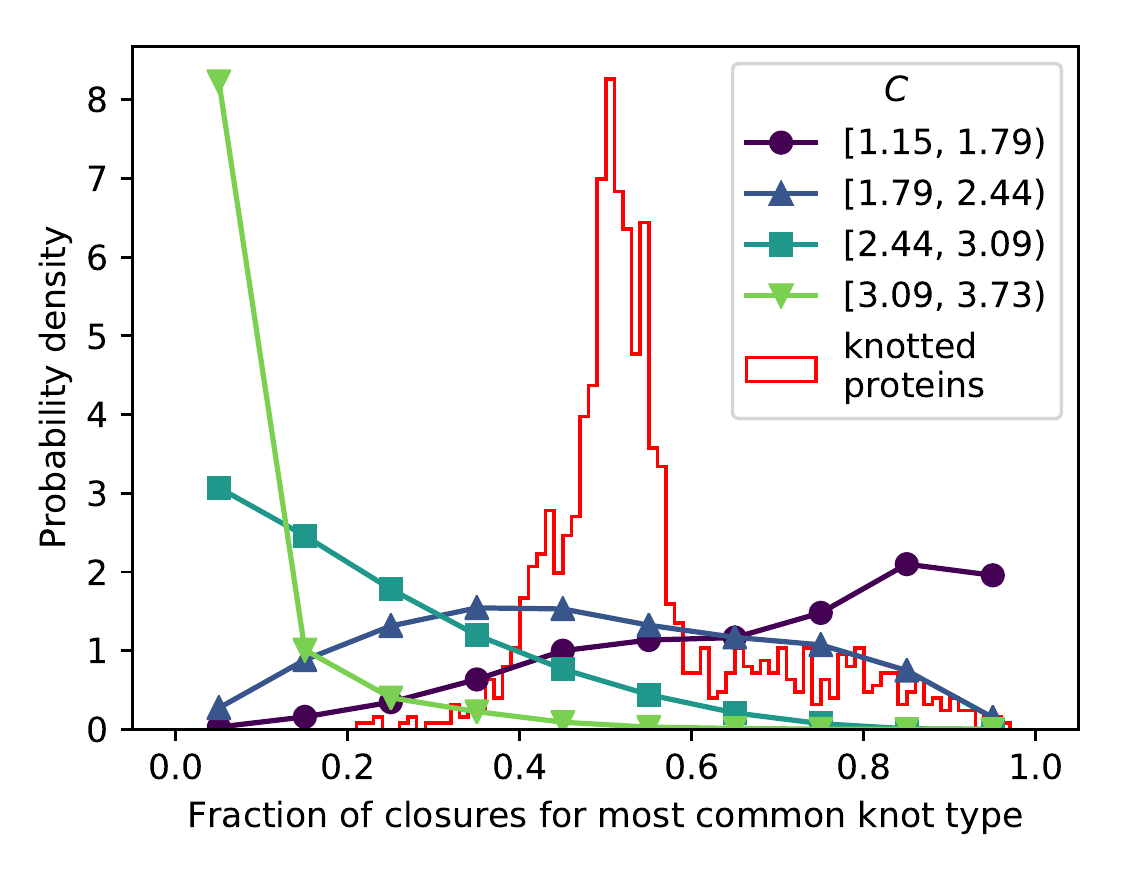}
 \caption{Probability distribution for the coverage of the most common knot for knotted ideal chains, compared to all knotted proteins in the Protein Data Bank.
 }
 \label{proteins_comparison}
\end{figure}

The original motivation of our initial investigation of virtual closure analysis of open chains was to study knotted proteins~\cite{alexander17}, which are examples of macromolecules which are compact and not completely flexible. 
Our results revealed that, while proteins in the Protein Data Bank are rarely knotted, almost 40\% are weakly knotted.
The combination of a low knotting probability and a relatively high proportion of weak knotting is unlike any of the random walk models investigated here.
Fig.~\ref{proteins_comparison} provides more detail, showing a comparison between the distribution of most common knot coverage in knotted proteins versus knotted, spherically confined, off-lattice walks.
Evidently, the protein case is very different from these walks, indicating the subtlety in modelling protein conformations as random chains.
The distribution is relatively sharp, with a peak in most common knot coverage only slightly above 50\%.
If a method even more sensitive to weak knotting were used such as knotoids considered on the plane~\cite{goundaroulis17b}, now implemented by the knotted proteins database KnotProt~\cite{knotprot}, it would not be surprising to find most knotted proteins classified as weakly knotted.
Much of the shape of this distribution can be attributed to the large number of knotted carbonic anhydrase structures which dominate the data set and are notable for the shallowness of their knotting.
As it is highly likely that a shallowly knotted chain is also weakly knotted, this accounts in part for the large number of protein chains with low fractional coverage for their commonest knot.

The distinction between weak and strong knotting is likely to be useful in the analysis of open chain dynamics. 
For instance, during the process of knot formation, a flexible open chain passes through one or more weakly knotted conformations before reaching a final deeply knotted state.
Similar transitions should be observable in unknotting pathways.
In these situations, the weakly knotted state is likely to be unstable, quickly untying or securing a knot with some perturbation to the endpoints.
In controlling chain dynamics, confinements or effective sizes might be manipulated to enhance the chance of a weakly knotted configuration, which might then make a further transition to a strongly knotted configuration.
Further study of the weak-strong transitions may reveal further structure in these distinctions of knot types for open curves.
An important extension to our work in this regard would be to investigate walks in tighter tubes and slits.
In such extremely confined environments, the geometrical complexity of the walks is likely to be reduced, resulting in a falling proportion of weak knots which we predict to be captured by a decreasing adjusted degree of confinement (defined in (\ref{eq:Cstardef})) as confining radius is decreased.
This is in line with the non-monotonic knotting behaviour previously seen~\cite{orlandini13}.
This work cannot be fully carried out using the algorithms for generating walks described here, however, as these require a radius of at least 1.

\appendix

\section{Generating ideal chains in spheres, tubes and slits}
\label{sec:appendix}

To generate ideal chains confined to the sphere, we follow the algorithm of Diao et al.~\cite{diao12} to ensure that walk vertices are uniformly distributed throughout the volume.
Here we summarise this model, with $R$ representing the radius of the confining sphere centred at the origin, and $X$ the previous endpoint of the walk whose next step is being considered, a distance $r = |X|$ from the centre of the sphere.
If $X$ is further than a unit from the spherical boundary, the next point is chosen by stepping a unit in a uniformly random direction.
If $X$ is within a unit of the boundary, the next step's direction is sampled from a distribution weighted towards the boundary: this depends on the choice of an azimuthal angle $\phi$ and polar angle $\theta$ with respect to an axis given by the direction of $X$ with respect to the centre of the sphere. 
As the problem is axisymmetric, $\phi$ is uniformly distributed between $0$ and $2\pi$.
The value of $-\cos \theta$ is sampled differently in each of the following three ranges: angles which, on a unit sphere about X, lie further than a unit from the boundary, angles within a unit of the boundary, and outside the boundary. 
The values of $-\cos \theta$ which lie within these categories are:
 \begin{equation} \fl
  \hbox{$-\cos \theta$ = } 
  \cases{
   \text{further than a unit from the boundary,}     &   $-1\phantom{(R)} \leq - \cos \theta \leq a(r,R)$; \cr
   \text{within a unit of the boundary,}    &    $a(r,R) < - \cos \theta \leq b(r,R)$; \cr
   \text{outside the boundary,}   &     $b(r,R) < - \cos \theta$, \cr
  }
  \label{diao_sphere_theta_boundaries}
 \end{equation}
where
\begin{equation} \fl
  a(r, R) = 
  \cases{
   \case{1}{2r}(R^2 - r^2 - 2R),    &    $r > R - 1 \text{ and } R - 1 > \min(r, |r-1|)$; \cr
   -1,    &    $r > R - 1 \text{ and } R - 1 \leq \min(r, |r - 1|)$; \cr
   1,    &    $r \leq R - 1$, \cr
  }
  \label{diao_sphere_a}
\end{equation}
marking the angle which would step a unit from the boundary, and
\begin{equation}
 b(r, R) = 
  \cases{
   \case{1}{2r}(R^2 - r^2 - 1),    &    $r > R - 1$; \cr 
   1,    &    $r \leq R - 1$, \cr
  }
  \label{diao_sphere_b}
\end{equation}
marking the angle which would step to the boundary.

The probability density function (PDF) proposed by \cite{diao12} from which $-\cos \theta$ is sampled is uniform for angles further than a unit from the boundary, rises linearly within a unit of the boundary, and is zero outside of the boundary.
Explicitly, this is 
\begin{equation} \fl
  \text{PDF}(-\cos \theta) = 
  \cases{
   \case{1}{2},    &    $-1\phantom{(R)} \leq - \cos \theta \leq a(r,R)$; \cr
   \case{1}{2}(1 + c[- \cos \theta - a(r,R)]),    &    $a(r,R) < - \cos \theta \leq b(r,R)$; \cr
   0,    &    $b(r,R) < - \cos \theta$ \cr 
  }
  \label{diao_sphere_theta_pdf}
\end{equation}
where 
\begin{equation}
  c = \frac{4r((r + 1)^2 - R^2)}{(2R - 1)^2}
  \label{diao_sphere_c}
 \end{equation}
for normalisation.
The radial distribution of vertices for walks generated in this way is shown in Fig.~\ref{radial_distribution}~(a), compared with both the desired uniform distribution and vertices for walks in absorbing boundaries.
  
We adapt this method to find distributions of vertices of chains confined in other geometries.
The case of confinement in slits, or between two parallel walls, is relatively straightforward by approximating each wall as a sphere with large radius, in practice 10,000.
The same PDF gives almost uniformly distributed vertices in the slit.

Confinement in a tube is more challenging since the problem is no longer axisymmetric in the same way as the sphere.
The following modification of the model of \cite{diao12} gives a distribution of vertices close to uniform, although not quite as good as the spherical case.
Align the tube along the $z$-axis, and take $\phi$ now as the polar angle with respect to this direction, with $\cos \phi$ chosen uniformly between $0$ and $\pi$.
$\theta$ is now the azimuthal angle, and it again determines how close the step is to the boundary, which also now depends on $\phi$.
Once $\phi$ is chosen, the next step can only lie on a section of a circle situated within the tube.
Unlike with spherical confinement, the choice of $\phi$ alters the radius of this circle, from 1 at $\phi=\pi/2$ to 0 at $\phi=0 \text{~or~} \pi$.
To account for this when sampling $\theta$, we scale the problem to give an effective $r$ of
\begin{equation}
  r_{\text{eff}}=r_{X}/\sin\phi,
  \label{diao_tube_effective_r}
\end{equation}
where $r_X$ is the actual radial distance of $X$, and an effective $R$ of  
 \begin{equation}
  R_{\text{eff}} = R_{\text{tube}} / \sin\phi,
  \label{diao_tube_effective_R}
\end{equation}
where $R_{\text{tube}}$ is the actual radius of the tube.

The tube-confined $\theta$ is sampled with respect to cylindrical radius between $0$ and $\pi$, clockwise or anticlockwise at random.
The PDF (\ref{diao_sphere_theta_pdf}), removing the cosine, overcompensates the boundary-avoiding behaviour.
In order to limit this rise in the region within a unit of the boundary to be slower than linear, we used the distribution
 \begin{equation}
  \text{PDF}(\theta) = 
  \cases{
   \case{1}{\pi},    &    $-\pi \leq \theta \leq \alpha$; \cr
   \case{1}{\pi}(1 + c(\theta - \alpha)^n),    &    $\alpha < \theta \leq \beta$; \cr
   0,    &    $\beta < \theta$ \cr
  }
  \label{diao_cylinder_theta_pdf}
 \end{equation}
 where the angles $\alpha =\arccos(-a)$ and $\beta = \arccos(-b)$ are equivalent to the boundaries used in the case of the sphere case, and 
 \begin{equation}
  c =  -\frac{\beta(n + 1)}{(\beta - \alpha)^{n+1}}
  \label{diao_cylinder_c}
 \end{equation}
for normalisation.
The resulting radial distribution of walk vertices using this corrected PDF is plotted in Fig.~\ref{radial_distribution}~(b) together with the distribution using an absorbing boundary, and the original spherical PDF.
This used a value of $n=0.6$, which we found gave the closet fit to uniform.
Evidently, the fit is an improvement, although not ideal.
The small deviation is unlikely to have a large effect on the knotting statistics.

\ack
We are grateful for conversations with Benjamin Bode, Enzo Orlandini and Stu Whittington.
In particular, we thank Alyona Akimova Andreevna for correspondence on genus one virtual knots.
This research was supported by the Leverhulme Trust Research Programme RP2013-K-009, SPOCK: Scientific Properties of Complex Knots.

\end{document}